\documentclass{ws-procs9x6}

\def\etal {{\it et al.}}

\begin{document}

\title{THE KATRIN EXPERIMENT: STATUS AND OUTLOOK}

\author{D.S.\ PARNO}

\address{Center for Experimental Nuclear Physics and Astrophysics\\
and Department of Physics, University of Washington\\
Seattle, WA 98195, USA\\
E-mail: dparno@uw.edu}

\author{On behalf of the KATRIN Collaboration}

\begin{abstract}
The KATRIN experiment, presently under construction in Karlsruhe, Germany, will improve on previous laboratory limits on the neutrino mass by a factor of ten. KATRIN will use a high-activity, gaseous T$_2$ source and a very high-resolution spectrometer to measure the shape of the high-energy tail of the tritium-decay $\beta$ spectrum. The shape measurement will also be sensitive to new physics, including sterile neutrinos and Lorentz violation. This report summarizes recent progress in the experiment.
\end{abstract}

\bodymatter

\phantom{x}
\vskip 10pt

\noindent
Neutrino oscillation experiments show that the neutrino-mass eigenstates $\nu_1$, $\nu_2$, $\nu_3$ are distinct from the weak-flavor eigenstates $\nu_e$, $\nu_\mu$, $\nu_\tau$.~\cite{pdg12} In the quasi-degenerate regime, where the mass splittings are small compared to the absolute mass scale, laboratory $\beta$-decay experiments can extract the antineutrino mass through a shape analysis of the high-energy tail of the $\beta$ spectrum, making no model assumptions apart from basic kinematics. Previous efforts to perform such a measurement using tritium have yielded an upper limit of $\sqrt{\sum_i |U_{ei}|^2 m_{\bar{\nu}_i}^2} < 2 \mathrm{eV}$ at 90\% confidence, where $m_{\bar{\nu}_i}$ is the mass of the $i$th neutrino-mass eigenstate and $U_{ei}$ is an element of the neutrino mixing matrix.~\cite{pdg12} The KArlsruhe TRItium Neutrino (KATRIN) experiment will improve this sensitivity by a factor of 10.~\cite{katrincdr}

\begin{figure}[tb]
\begin{center}
\psfig{file=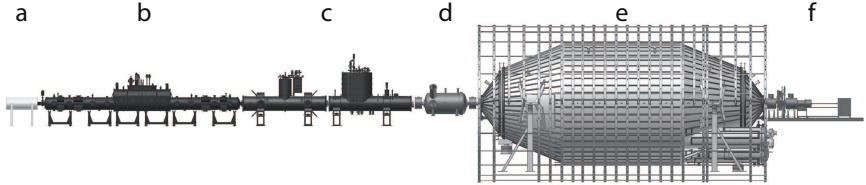,width=4.5in}
\end{center}
\caption{The KATRIN beamline.} 
\label{katrin_beamline}
\end{figure}

Figure~\ref{katrin_beamline} shows the KATRIN beamline. The rear section (a) monitors the windowless, gaseous T$_2$ source (WGTS, b), in which gaseous tritium decays in a magnetic field. A two-stage transport section (c) adiabatically guides $\beta$s along magnetic field lines toward a pair of spectrometers (d, e). Each spectrometer is a magnetic adiabatic collimating / electrostatic filter~\cite{macefilter}. As $\beta$s travel from a region of strong magnetic field to a field minimum, their magnetic orbital moments remain constant while their momenta are rotated to be nearly longitudinal. A longitudinal retarding potential thus sets a kinetic-energy threshold for $\beta$s passing through to the detector (f). By adjusting the retarding potential and thus the integration threshold, KATRIN will make a precise measurement of the tritium $\beta$ spectrum near the endpoint. The fit to the spectral shape will have at least four free parameters: the antineutrino mass, the $\beta$ endpoint energy, the signal rate, and the background rate. 

Here, we highlight some recent achievements of the KATRIN collaboration, from research and development to construction and commissioning.

The 10-m source tube, presently under construction, will use components of the WGTS Demonstrator apparatus, a prototype used to test aspects of the source design. Using a cooling system based on neon thermosiphons, the Demonstrator was maintained for a week at the design temperature of 30~K. The demonstrated temperature stability of $\Delta T/T = 5 \times 10^{-5}$~ h$^{-1}$ is a factor of 20 better than the design specification.~\cite{wgts_demonstrator_tempcontrol}

The molecular composition of the source affects the source activity as well as the molecular final state after decay of a T atom. The LAser RAman (LARA) system uses Raman spectroscopy to continuously monitor the composition of the gas to be injected into the source. In a prototype of the circulation system, 0.1\% precision was demonstrated in 250~s of acquisition.~\cite{fischer_lara}

The commissioning of the pre-spectrometer revealed potential pitfalls in the spectrometer design. For example, ring-shaped background patterns were observed to accumulate over the course of several hours. Through analysis, simulation, and dedicated runs, these were traced to radon atoms that escaped from the getter pumps and decayed inside the spectrometer flux tube. A shakeoff or internal-conversion electron from such a decay could be stored in the spectrometer for hours, slowly losing energy via collisions with residual gas molecules. The resulting low-energy secondary electrons were then accelerated by the retarding potential; their distribution on the detector plane was ring-shaped due to the magnetron motion of the primary.~\cite{fraenkle_radon, radon_modeling, radon_validation} With the identification of this new background source, the main-spectrometer design was modified to include liquid-nitrogen-cooled, radon-trapping baffles between the getter pumps and the main volume.

A two-layer wire electrode system containing 23,440~wires lines the 700-m$^2$ interior surface of the main-spectrometer vessel, which is 23.8~m long with a maximum diameter of 9.8~m.~\cite{wire_electrode}. This system improves field stability and allows fine shaping of the retarding potential. Since the wire electrodes are held at a slightly more negative potential than the hull of the vessel, they also reduce background by preventing secondary electrons from entering the flux tube. This system has been installed and baked out to a temperature of 300$^\circ$C.

The 0.93-eV design resolution of the main spectrometer is set by the ratio of the minimum magnetic field, in the central analyzing plane, to the maximum magnetic field at one end. In order to achieve a magnetic-field minimum of precisely 3~G, two sets of air coils have been installed around the main spectrometer.~\cite{aircoil_emd} The circular low-field-correction coils, which are coaxial with the main spectrometer, compensate for distortions due to residual magnetization and to fringe fields from the superconducting solenoids at either end. Compensation for the Earth's magnetic field is achieved by coils in straight sections, describing a cylindrical surface centered on the flux tube and joined by vertical endrings. The resulting magnetic field is monitored by stationary sensors and by robotic mobile sensor units that travel along the support structures for the low-field-correction coils.~\cite{mobs}

The focal-plane detector is a monolithic silicon PIN diode; its active area has a diameter of 90~mm. It is segmented into 148~pixels laid out in a dartboard pattern, allowing the characterization of inhomogeneities in the analyzing plane. A novel mounting structure~\cite{carac}, in which electrical contacts are made by spring-loaded pogo pins, minimizes radioactivity in the region of the detector. With a built-in calibration source, we have demonstrated an energy resolution of $1.48 \pm 0.01$~keV (full width at half maximum) for electrons at the tritium $\beta$-spectrum endpoint energy of 18.6~keV.

In May 2013, the main spectrometer and detector sections were connected for the first time, beginning an initial commissioning period to study backgrounds and the spectrometer transmission function. After hardware upgrades, a second commissioning phase will follow in 2014, along with commissioning of the source and transport sections. We anticipate the start of KATRIN data-taking in late 2015.  

KATRIN expects to achieve a sensitivity of $\sqrt{\sum_i |U_{ei}|^2 m_{\bar{\nu}_i}^2} < 0.2$~eV at the 90\% confidence level in three full years of running, or approximately five calendar years. The precise spectral-shape analysis also gives potential access to new physics, including sterile neutrinos in the mass range suggested by the reactor antineutrino anomaly.~\cite{sterile_nu} Endpoint measurements from previous tritium-based neutrino-mass experiments have recently been used to set limits on the coefficients of two Lorentz-violating operators in the Standard-Model Extension, which KATRIN could improve by two orders of magnitude~\cite{lorentz_violation}. The KATRIN measurement plan, which involves cycling through all retarding-potential measurements with a period of about 15~minutes~\cite{bayesian_analysis}, will give access to the complex coefficient of an additional Lorentz-violating operator that produces oscillations of the endpoint with a period of one sidereal day.~\cite{lorentz_violation}

\section*{Acknowledgments}

The KATRIN experiment is funded by the Helmholtz Gemeinschaft, the Bundesministerium f\"ur Bildung und Forschung, and the US Department of Energy. The author is funded by grant number DE-FG02-97ER41020 from the US Department of Energy.


\begin{thebibliography}{xx}

\bibitem{pdg12}
Particle Data Group,
J.\ Beringer \etal,
Phys.\ Rev.\ D {\bf 86}, 010001 (2012).

\bibitem{katrincdr}
J.\ Angrik \etal,
{\it KATRIN Design Report 2004,}
preprint FZKA-7090 (2005).

\bibitem{macefilter}
G.\ Beamson, H.Q.\ Porter, and D.W.\ Turner,
J.\ Phys.\ E: Sci.\ Instrum.\ {\bf 13}, 64 (1980).

\bibitem{wgts_demonstrator_tempcontrol}
S.\ Grohmann \etal,
Cryogenics {\bf 55-56}, 5 (2013).

\bibitem{fischer_lara}
S.\ Fischer \etal,
Fusion Sci.\ Technol.\ {\bf 60}, 925 (2011).

\bibitem{fraenkle_radon}
F.M.\ Fr{\"a}nkle \etal,
Astropart.\ Phys.\ {\bf 35}, 128 (2011).

\bibitem{radon_modeling}
N.\ Wandkowsky \etal,
arXiv:1304.1375.

\bibitem{radon_validation}
N.\ Wandkowsky \etal,
J.\ Phys.\ G
{\bf 40}, 085102 (2013).

\bibitem{wire_electrode}
K.\ Valerius,
Prog.\ Part.\ Nucl.\ Phys.\ {\bf 64}, 291 (2010).

\bibitem{aircoil_emd}
F.\ Gl{\"u}ck \etal,
arXiv:1304.6569.

\bibitem{mobs}
A.\ Osipowicz \etal,
J.\ Inst.\ {\bf 7}, T06002 (2012).

\bibitem{carac}
B.A.\ VanDevender \etal,
Nucl.\ Inst.\ Meth.\ Phys.\ Res.\ A {\bf 673}, 46 (2012).

\bibitem{sterile_nu}
J.\ Formaggio and J.\ Barrett,
Phys.\ Lett.\ B {\bf 706}, 68 (2011).

\bibitem{lorentz_violation}
J.S.\ D{\'i}az, V.A.\ Kosteleck{\'y}, and R.\ Lehnert,
arXiv:1305.4636.

\bibitem{bayesian_analysis}
A.\ Sejersen Riis, S. Hannestad, and C.\ Weinheimer,
Phys.\ Rev.\ C {\bf 84}, 045503 (2011).

\end{thebibliography}
\end{document}